\documentclass[%
reprint,
groupedaddress,
amsmath,amssymb,
aps,prl,
twocolumn,
showpacs,
]{revtex4-1}
\usepackage{graphicx}
\usepackage{dcolumn}
\usepackage{bm}
\usepackage{braket}
\usepackage{hyperref}
\usepackage[caption=false]{subfig}
\usepackage{amsmath}
\usepackage{fancyhdr}
\usepackage{float}
\usepackage{siunitx}
\usepackage{xcolor}
\usepackage{xspace}
\usepackage{eucal}
\usepackage{multirow}
\usepackage[utf8]{inputenc}

\newcommand{\Rb}{\textsuperscript{87}Rb\xspace}

\usepackage{ulem}

\begin{document}

\newcommand{\rev}[2]{#2}

\title{
Optomechanical resonator-enhanced atom interferometry 
}
\author{L.~L.~Richardson}
\altaffiliation{Current address: College of Optical Sciences, University of Arizona, Tucson, AZ 85721, USA}
\author{A.~Rajagopalan}
\author{H.~Albers}
\author{C.~Meiners}
\author{D.~Nath}
\author{C.~Schubert}
\altaffiliation{Current address: Deutsches Zentrum f\"ur Luft- und Raumfahrt e.V. (DLR), Institut für Satellitengeod\"asie und Inertialsensorik, c/o Leibniz Universit\"at Hannover, DLR-SI, Callinstraße 36, 30167 Hannover, Germany}
\author{D.~Tell}
\author{\'E.~Wodey}
\author{S.~Abend}
\author{M.~Gersemann}
\author{W.~Ertmer}
\author{E.~M.~Rasel}
\author{D.~Schlippert}\email[Electronic mail: ]{schlippert@iqo.uni-hannover.de}
\affiliation{Leibniz Universit\"at Hannover, Institut f\"ur Quantenoptik,\\ Welfengarten 1, 30167 Hannover, Germany}
\author{M.~Mehmet}
\affiliation{Institut f\"ur Gravitationsphysik / Albert-Einstein-Institut~(AEI), Leibniz Universit\"at Hannover,
Callinstra\ss e 38, 30167 Hannover, Germany}
\author{L.~Kumanchik}
\author{L.~Colmenero}
\author{R.~Spannagel}
\author{C.~Braxmaier}
\affiliation{German Aerospace Center~(DLR) -- Institute of Space Systems \& University of Bremen -- Center of Applied Space Technology and Microgravity~(ZARM), Robert-Hooke-Straße 7, 28359 Bremen, Germany}

\author{F.~Guzm\'an}\email[Electronic mail: ]{guzman@tamu.edu}
\altaffiliation{Current address: Texas A\&M University, Aerospace Engineering \& Physics, College Station, TX 77843, USA}
\affiliation{College of Optical Sciences, University of Arizona, Tucson, AZ 85721, USA,
\\German Aerospace Center~(DLR) -- Institute of Space Systems \& University of Bremen -- Center of Applied Space Technology and Microgravity~(ZARM), Robert-Hooke-Straße 7, 28359 Bremen, Germany}
\date{\today}

\begin{abstract}
Matter-wave interferometry and spectroscopy of optomechanical resonators \rev{not only belong to the most successful methods in quantum optics, but also}{} offer complementary advantages.
Interferometry with cold atoms is employed for accurate and long-term stable measurements, yet it is challenged by its dynamic range and cyclic acquisition. 
Spectroscopy of optomechanical resonators features continuous signals with large dynamic range, however it is generally subject to drifts. 
In this work, we combine the advantages of both \rev{quantum-optical}{} devices. 
Measuring the motion of a mirror and matter waves interferometrically with respect to a joint reference allows us to operate an atomic gravimeter in a seismically noisy environment otherwise inhibiting readout of its phase.
Our method is applicable to a variety of quantum sensors and shows large potential for improvements of both elements by quantum engineering.
\end{abstract}
\maketitle
\begin{figure*}[t]
\begin{center}
\includegraphics[width=\linewidth]{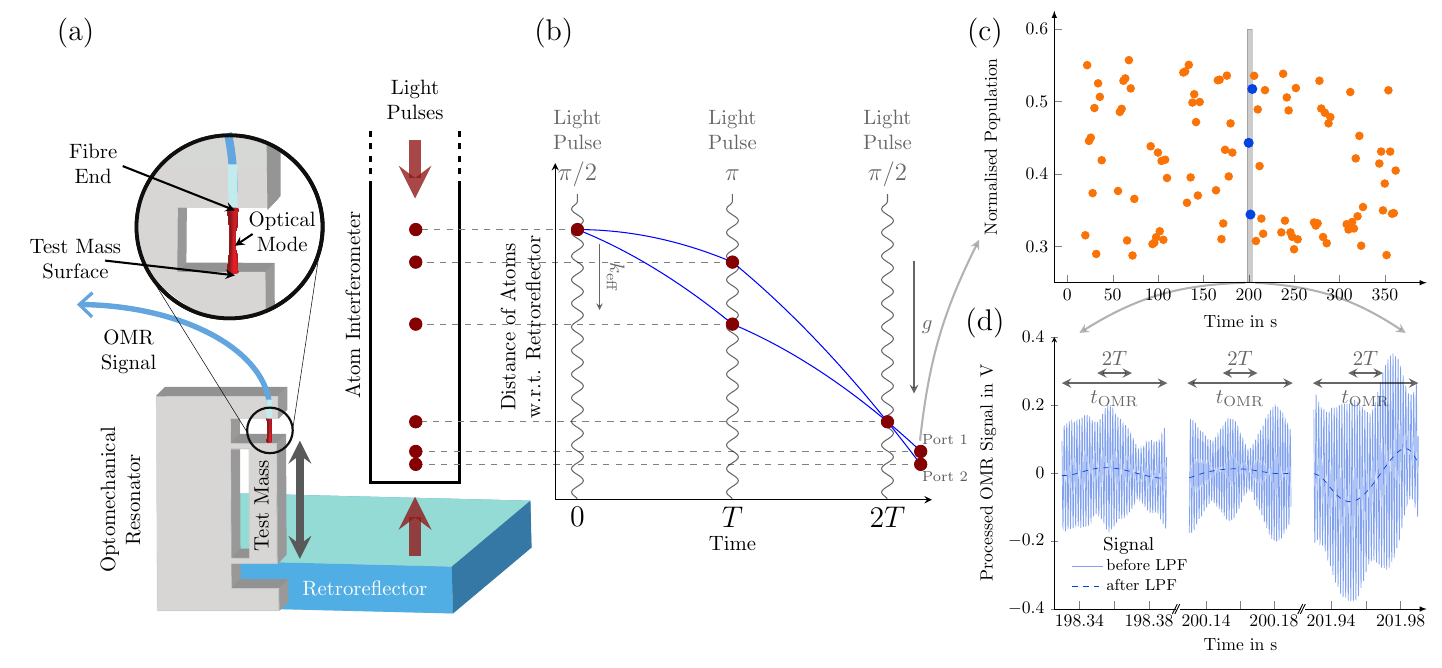}
\caption{(a)~Schematic of the experimental setup (not to scale) comprising an optomechanical resonator~(OMR) enhancing the optical mode between the flat end of a polarisation-maintaining fibre (light blue) and a side face of the test mass as detailed in the enlarged view and a Mach-Zehnder-type atom interferometer measuring the gravitational acceleration $g$ and (b)~its spacetime diagram. 
The sensor is attached to a retroreflection mirror which rests in ``strapdown'' configuration on a platform on the laboratory floor. 
(c)~Ambient vibrations drive the retroreflector's motion beyond the reciprocal range of the atom interferometer and thus obscure the correspondence between phase and population in the ports 1 and 2.
An interferometric fringe can be restored by measuring (d)~the test mass motion of the optomechanical  resonator and digitally convolving with the atom interferometer's acceleration response function.
Measurement intervals of both devices of durations $t_{\textnormal{OMR}}$ and $2T$ are synchronised~(details in the Methods section).
Typical OMR signals are shown before (blue solid line) and after (blue dashed line) low-pass filtering~(LPF) which removes the dominant mechanical resonance at \SI{678.5}{\hertz}.}
\label{img:setup}
\end{center}
\end{figure*}
Matter-wave interferometers \rev{have become a competitive tool}{are employed} in fundamental physics~\cite{Fixler07Science,RosiNature2014,Spagnolli17PRL,Fray04PRL,Bonnin13PRA,SchlippertPRL2014,Tarallo2014PRL,Zhou15PRL,KovachyNature2015,Pandey19Nature,HamiltonPRL2015,JaffeNaturePhys2017,Bouchendira2011PRL,Decamps16PRL,ParkerScience2018,Shayeghi20NatCommun}, metrology, and inertial sensing~\cite{PetersNature1999,HuPRA2013,FangJOP2016,FreierJOP2016,Hardman16PRL,MenoretSR2018,Bidel18NatComm,Stockton2011PRL,BergPRL2015,DuttaPRL2016,Hinton17PhilTrans,Chen20PRL}.
In recent years, developments in the quantum engineering of optomechanical resonators have yielded devices with exciting applications in fields such as quantum information, fundamental physics and, likewise, in inertial sensing~\cite{GuzmanAPL2014,Pisco2018Nature}.
In addition, cold atoms were coupled to micromechanical cantilevers~\cite{HungerPRL2010} and nanomembranes~\cite{CamererPRL2011}.
\rev{Here we combine an optomechanical resonator with a light-pulse atom interferometer and measure the accelerations of the resonator's test mass and a freely falling cloud of atoms relative to a joint reference, the atom interferometer's retroreflection mirror~(Fig.~\ref{img:setup}~(a)). By merging the complementary benefits of both sensors, this arrangement allows us to restore the interference fringes of the atomic interferometer and thus to operate the atom interferometer under strong seismic perturbations without loss of phase information.}{}
So far, commercial \rev{motion}{} sensors have been exploited \rev{}{to track the motion of atom interferometers' inertial references,} \rev{for dynamic-range extension, vibration-noise suppression~\cite{KohelESTC2008,LeGouetAPB2008,MerletMetrologia2009,LautierAPL2014,GeigerNatureComm2011,BarrettNatureComm2016, Cheiney2018}, as well as comparisons with atom interferometers~\cite{peters2001,Merlet2010,FreierJOP2016}.}{thus extending the measurement dynamic range, suppressing vibration noise~\cite{KohelESTC2008,LeGouetAPB2008,MerletMetrologia2009,LautierAPL2014,GeigerNatureComm2011,BarrettNatureComm2016, Cheiney2018}, and allowing for comparisons with other atom interferometers~\cite{peters2001,Merlet2010,FreierJOP2016}.}

\rev{}{Here we combine a high-bandwidth optomechanical resonator with a long-term stable light-pulse atom interferometer and measure the accelerations of the resonator's test mass and a freely falling cloud of atoms relative to the atom interferometer's inertial reference.}
\rev{}{Our atom interferometer measures gravity under strong seismic perturbations without loss of phase information.}
\rev{In contrast, our method merges two quantum-optical systems into one unit. Both systems benefit from the large toolbox of quantum engineering usually exploited in optical spectroscopy and photonics and are highly customisable.}{In contrast to previous approaches, our method merges two systems both benefiting from the large toolbox of quantum engineering usually exploited in optical spectroscopy and photonics to a highly customisable device.}
To this end, extensions of our method will make use of the tunability of the mechanical resonance of optomechanical resonators.
In addition, \rev{the combination of}{} devices with different resonance frequencies or different orientations will grant access to larger bandwidth and multiple sensitive axes.
In combination with the use of more advanced spectroscopy and feedback methods this opens up atom interferometry to rough environments.
\rev{Indeed, applications of our}{Our} method \rev{are}{is} not restricted to gravimetry or navigation~\cite{JekeliNav2005} but can help to enhance any interferometric measurement with matter as well as with light~\cite{Niebauer95Metrologia} in environments with large inertial noise, thus replacing bulky vibration isolation and motion sensors.
\rev{This might also apply when one combines our optomechanical resonator for compensating inertial noise in the resonators of optical clocks, e.g., in transportable setups~\cite{Grotti18NatPhys}. In that respect, our method shares analogies with atomic clocks by hybridisation of long- and short-term references.}{}
Finally, optomechanical resonators with a volume of much less than a cubic centimetre \rev{not only allow measuring precisely at the point of interest, but also}{} show a great potential for miniaturisation \rev{}{of future quantum sensors}.
\section{Results}
\begin{figure}[t]
\begin{center}
\includegraphics[width=\linewidth]{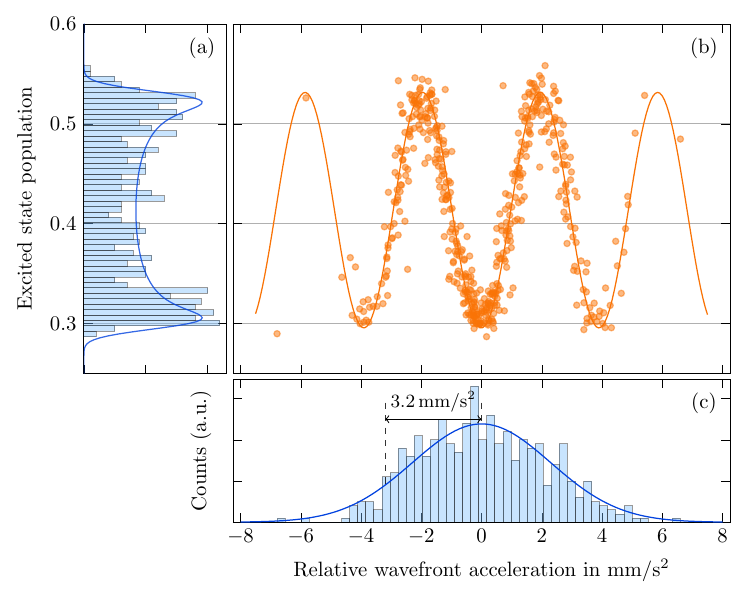}
\end{center}
\caption{(a)~Histogram distribution of the normalised interferometer output ports, (b)~a fringe (orange circles) recovered by post-correction based on the resonator signal, and a sinusoidal fit to the corrected data~(orange solid line) for a pulse separation time $T=\SI{10}{ms}$. 
(c)~Ambient vibrations with a Gaussian $1/e$ width of \SI{3.2}{mm/s^2}, if uncorrected, obscure the phase information of the atom interferometer.
By convolving the recorded time series of the resonator signal with the interferometer acceleration response function we can recover each data point's phase information and reconstruct the fringe pattern.
Each histogram and the corresponding interferometer response represent a segment of 500 data points.
}
\label{img:post-correction}
\end{figure}

We operate a Kasevich-Chu interferometer~\cite{KasevichPRL1991} combined with an optomechanical resonator attached to the mirror providing the phase reference for the interferometer as a novel gravimeter (\rev{Fig.~\ref{img:setup}~(b)}{Fig.~\ref{img:setup}~(a) \& (b)} and details in the Methods section).
Ambient vibrational noise couples to the retroreflector at a weighted acceleration level of {\SI{3}{mm/s^2}} per cycle.
This leads to phase excursions exceeding a single fringe during one interferometric measurement with the readout appearing to be random due to the underlying $2\pi$ phase ambiguity~(Fig.~\ref{img:setup}~(c)). 
Accordingly, the atom interferometer signal, i.e. the relative population of the two ports, features a bimodal distribution visible in Fig.~\ref{img:post-correction}~(a). 
However, ambient vibrational noise also results in a displacement of the resonator test mass which is recorded by the signal retroreflected from the optomechanical resonator~(Fig.~\ref{img:setup}~(d)).
The records of the optomechanical resonator make it possible to reconstruct the atomic interference pattern (Fig.~\ref{img:post-correction}~(b)). 
The signal from the optomechanical resonator is constrained to the band of interest.
We apply high-pass filters at \SI{0.8}{Hz} to suppress low-frequency drifts, as well as a digital low-pass filter at \SI{50}{Hz}, the atom interferometer's corner frequency~(see Methods section).
We subsequently sample it digitally over {$\SI{60}{ms}$} centred around the central light pulse of each interferometer cycle.
The phase correction is finally calculated from the signal utilising the acceleration sensitivity function describing the atom interferometer's phase response~\cite{CheinetIEEE2008,Bonnin2015}. 
Residual systematic biases can be experimentally analysed as shown for commercial sensors in Ref.~\cite{MerletMetrologia2009}. 

\begin{figure}[t]
\begin{center}
\includegraphics[width=\linewidth]{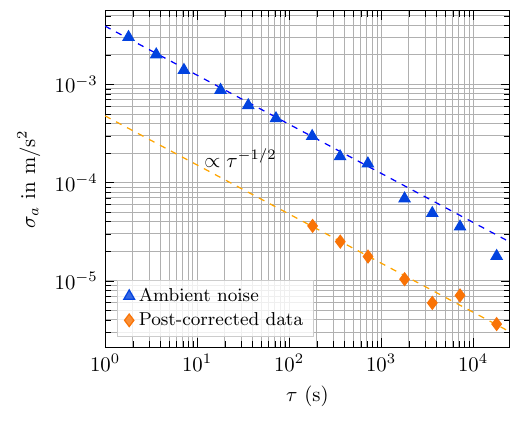}
\caption{Allan deviation of the measured gravitational acceleration as a function of integration time for estimated ambient noise~(blue triangles) and post-corrected data~(orange diamonds). 
Post correction improves the instability at \SI{1}{\second} by a factor of 8 and, here, reduces the measurement time necessary to achieve a target instability by a factor 64.
The dashed lines reflect the gain by averaging.
We estimate the ambient noise from the phase correction data obtained in our post correction (see Methods section).}
\label{img:ADEV}
\end{center}
\end{figure}

Using our method, we measure the local gravitational acceleration $g$ in an approximately \SI{22}{h}-long, interruption-free, measurement series otherwise impossible when operating both sensors alone. 
By suppressing vibrational noise, our sensor fusion method improves the overall short-term stability by a factor 8~(Fig.~\ref{img:ADEV}).
\begin{figure}[t]
\begin{center}
\includegraphics[width=\linewidth]{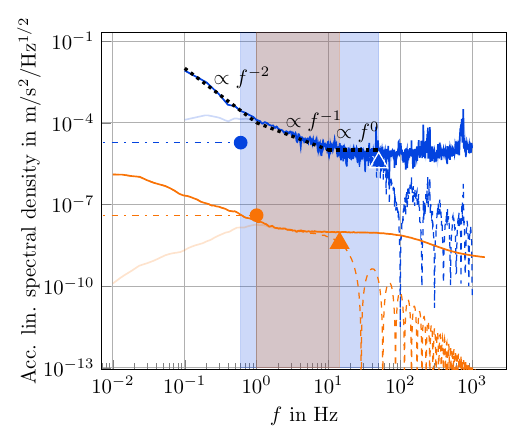}
\caption{Current and anticipated sensitivity of an atom interferometer enhanced by an optomechanical resonator. 
Current \rev{}{intrinsic} performance (blue dash-dotted line) is achieved using a $T = \SI{10}{\milli\second}$ interferometer with cycle frequency $f_c$  = \SI{0.6}{Hz}, momentum separation of two photon recoils and residual \rev{phase}{technical} noise of \SI{30}{mrad}. 
It can be improved with a pulse separation time $T = \SI{35}{\milli\second}$, a cycle frequency $f_c$  = \SI{1}{\hertz}, momentum separation of 8 photon recoils and residual phase noise \SI{3}{mrad} (orange dash-dotted line). 
Similarly, current optomechanical resonator acceleration sensitivity at quiet conditions (solid blue curve) with a resonance of \SI{678.5}{\hertz} and optical finesse 2 can be optimised by choosing a resonance frequency of \SI{1500}{Hz} and finesse of 1600~(solid orange curve). 
Dashed lines indicate the intrinsic noise of the optomechanical resonator weighted by the sensitivity function of the atom interferometer \rev{}{and high-pass filtered at \SI{0.8}{Hz} (\SI{1}{Hz}) for the current (advanced) scenario to suppress additional noise in the low-frequency band}.
The shaded areas bounded by $f_c$ (disks) and the atom interferometer’s corner frequency (triangles) mark the respective dominant frequency bands most relevant for optimal post-correction of seismic noise and motivate high-pass filtering the resonator signal (light blue and orange solid lines).}
\label{img:Sa_comparison}
\end{center}
\end{figure}
\section{Discussion}
Fig.~\ref{img:Sa_comparison} illustrates the present and projected features of the atom interferometer, the optomechanical resonator, and the combination of both.
We expect a large potential for improvements for both quantum-optical devices.
Our optomechanical resonator features a sensitivity comparable to commercial accelerometers.
Its sensor fusion performance is nevertheless limited by low frequency noise~(Fig.~\ref{img:Sa_comparison}, solid blue trace).
It displays a RMS white acceleration noise of $\SI{1E-5}{m/s^2/\sqrt{Hz}}$ between 10 and \SI{50}{\hertz}. 
Pink noise ($\propto f^{-1}$) dominates from 1 to \SI{10}{\hertz}. 
Below \SI{1}{\hertz}, Brownian noise ($\propto f^{-2}$) processes mainly caused by the optical fibre employed for interrogating the resonator prevail.
Since the resonator's sensitivity to accelerations increases quadratically with decreasing mechanical resonance frequency and linearly with optical finesse, there is room for improvements by trading sensitivity against larger bandwidth and dynamic range~\footnote{Increasing the sensitivity through lower natural frequency of the resonator, the dynamic range is necessarily reduced since the device is operated in the non-linear regime.}. 
We foresee an optimisation of the hybrid sensor~\rev{}{(Fig.~\ref{img:Sa_comparison}, solid orange trace)} by \rev{designing}{tuning} the \rev{optomechanical resonator with a}{} resonance frequency \rev{of}{to} \SI{1500}{Hz} \rev{}{to increase the bandwidth}, \rev{and}{} improving \rev{}{the optical finesse to 1600 by high-reflectivity coating~\cite{GuzmanAPL2014},} and the readout by an order of magnitude as compared to Ref.~\cite{GuzmanAPL2014} \rev{}{by means of spectroscopy techniques developed for ultrastable resonators, e.g., Pound-Drever-Hall locking~\cite{Pound,Drever}}. 
Millimetre-sized optomechanical resonators have \rev{}{already} demonstrated sensitivities of \SI{1e-6}{m/s^2/\sqrt{Hz}} over bandwidths up to \SI{12}{kHz}.\rev{ and spectroscopy techniques developed for ultrastable resonators~\cite{Pound,Drever} can be exploited to improve the performance of the readout.}{}

Additionally, pathways exist for future atom interferometers customised for gravimetry, as discussed in Ref.~\cite{AbendPRL2016}. 
The sensitivity can be enhanced by operating the device with $T=\SI{35}{ms}$, a cycle rate of \SI{1}{Hz}, higher-order Bragg processes transferring $4 \cdot \vec{k}_\textnormal{eff} $ and a reduced phase noise of \SI{3}{mrad}. 
By improving the atom interferometer and tuning the optomechanical resonator, it is plausible that the intrinsic noise can be lowered to $\SI{6E-8}{m/s^2/\sqrt{Hz}}$. 
This target performance is comparable to the noise obtained in a quiet environment with an active vibration isolation~\cite{HuPRA2013} and outperforms transportable, commercial devices~\cite{MenoretSR2018}. 
This is based on the assumption of an environment described by the \rev{Peterson new high noise model}{``Peterson new high noise model''}~\cite{Peterson1993}.

Many atomic gravimeters employ rubidium and generate the light for manipulating the atoms by second harmonic generation from fibre lasers in the telecom C-band~\cite{QIN2019,THERON2017152}.
The inclusion of the optomechanical resonator therefore requires only minor hardware changes and can be performed with an all-fibred setup. 
\rev{The resonator shows improved performance in vacuum}{The resonator can be implemented directly into  inertial reference mirrors under vacuum, thus improving the mechanical quality factor while supporting miniaturisation of the overall setup.  It }does not emit notable heat, and is non-magnetic. \rev{ and consequently does not induce related errors~\cite{LeGouetAPB2008,Hu17PRA,HaslingerNaturePhys2017}}{Consequently, it does not induce systematic errors due to black body radiation~\cite{HaslingerNaturePhys2017} or due to spurious magnetic fields coupling to the matter waves~\cite{LeGouetAPB2008,Hu17PRA}, and neither do external magnetic fields couple to the resonator test mass.}
Moreover it can be easily merged with the retroreflection mirror of the atom interferometer.
Last but not least, the small volume of cubic millimetres offers great prospects for being integrated on atom chip sensors~\cite{AbendPRL2016}, and, hence, a large potential for miniaturisation of the sensor head.

\rev{}{Our method shares analogies with atomic clocks by hybridisation of long- and short-term references. Beyond this, our optical sensor might be used for compensating inertial noise in the resonators of optical clocks, e.g., in transportable setups~\cite{Grotti18NatPhys}.}

In conclusion, we have demonstrated an atom interferometer enhanced by an optomechanical resonator. 
We show operation of the atom interferometer under circumstances otherwise impeding phase measurements. 
Inertial forces on the atoms and on the resonator mirror are measured to the same reference permitting a direct comparison and high common mode noise suppression in the differential signal. 
Our method is not restricted to atomic gravimeters and could be beneficial to nearly all atom interferometric sensors. 
In particular, the achievable large dynamic range opens up great perspectives for the use of atomic sensors for inertial navigation~\cite{Cheiney2018} and airborne gravimetry~\cite{Bidel20JoG}. 
Finally, a possible modification of our setup's topology would ensure that the atom-optics light field is reflected directly off a micromechanical test mass.
In future experiments, we envisage exciting research on coherent light-mediated coupling of matter waves and mechanical systems~\cite{CamererPRL2011,KargPRA2019} with pulsed instead of cw-interaction.
\section{Methods}
\paragraph{Atom Interferometer --}

Our setup (Fig.~\ref{img:setup}~(a) \& (b)), which was employed as a differential gravimeter in Ref.~\cite{SchlippertPRL2014,Albers2020arxiv}, comprises a Kasevich-Chu interferometer~\cite{KasevichPRL1991}. 
In a $\pi/2 - \pi - \pi/2$ pulse sequence, stimulated two-photon Raman transitions coherently split, redirect, and recombine matter waves of \Rb.
The interferometer phase is determined by measuring the number of atoms in output ports 1 and 2 with state-selective fluorescence detection.
To leading order, a constant acceleration $\vec{a}$ of the atoms induces a phase shift 
\begin{equation*}
    \Delta\phi=\vec{k}_\textnormal{eff} \cdot \vec{a} \cdot T^2\, ,
\end{equation*}
where $\hbar\vec{k}_\textnormal{eff}$ is the photon recoil transferred to the atoms via a Raman process, and $T$ denotes the time between two subsequent light pulses. 
A chirp of the relative frequency of the lasers cancels the phase induced by acceleration, and is a measure for the latter. 
The atom interferometer's response to vibrational noise can be described using the sensitivity formalism~\cite{CheinetIEEE2008}.
Notably, the response is flat in a band between DC and up to the corner frequency $1/(2T)$, above which it features a low pass behaviour.
Typically, the interferometer’s response is adjusted by varying $T$ such that ambient noise induces phase shifts well within one fringe. 
At quiet conditions, i.e. with an operating vibration isolation system, the interferometer, which we operate at a cycle rate of {\SI{0.6}{Hz}}, features a fringe contrast of $\approx$ \SI{30}{\percent} and a Raman phase locked loop-limited phase noise of \SI{30}{mrad} for a time {$T=\SI{10}{\milli\second}$}.\\

\paragraph{Data Analysis --}
In order to suppress systematic shifts independent of the direction of momentum transfer we use the $k$-reversal method~\cite{McGuirk02PRA,Louchet-Chauvet11NJP}.
The interferometer measures 10 times in each direction of momentum transfer over a period of \SI{18}{s}.
For each scattering direction, we create histograms out of the normalised output population of the interferometer. 
Hereby each individual histogram comprises data accumulated over a {\SI{9000}{s}} period. 
From each histogram, we extract the interferometer response's amplitude and offset~\cite{VaroquauxNJP2009} as shown in Fig.~\ref{img:post-correction}. 
We subsequently extract $g$ for intervals of \SI{180}{s} (approximately 50 shots per direction) by fitting the post-corrected data to the expected atom interferometer response using the parameters determined by the histogram fit, solely leaving the interferometer phase as a free parameter. 
\rev{}{After \SI{18000}{s} integration we determine a local gravitational acceleration \SI{9.812675}{m/s^2} with a standard uncertainty of \SI{4e-6}{m/s^2}.}
We estimate the ambient noise in Fig.~\ref{img:ADEV} from the phase corrections made during post correction.
Accordingly, the data resembles the underlying ambient acceleration noise after weighting it with the atom interferometer's transfer function.
The first uncorrected value of $\approx \SI{3}{mm/s^2}$ per cycle also manifests in the $1/e$ width of the Gaussian-shaped spread in Fig.~\ref{img:post-correction}~(c).\\

\paragraph{Optomechanical Sensor --}
\begin{figure}[t]
\begin{center}
\includegraphics[width=\linewidth]{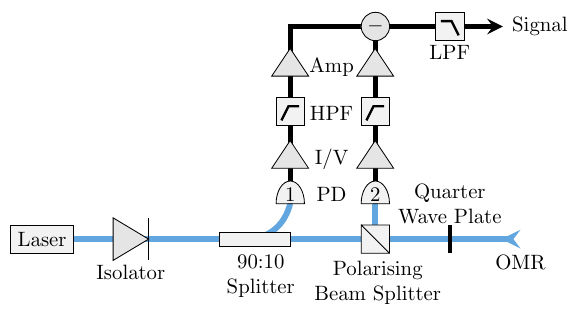}
\caption{Displacement readout system for the optomechanical resonator~(OMR).
Laser light split off for intensity noise correction and light reflected back from the OMR are detected on photo diodes~(PD 1 \& 2).
Both signals are passed to current-to-voltage converters~(I/V) followed by high-pass filters~(HPF) at \SI{0.8}{Hz} and are subsequently amplified.
Finally, the difference signal is digitally low-pass filtered~(LPF) at \SI{50}{Hz}.
}
\label{img:optics}
\end{center}
\end{figure}
Mirrors forming the optomechanical resonator, which has a volume on the order of a few hundred mm\textsuperscript{3}, are made from the flat tip of a polarisation maintaining fibre and a side face of the fused silica test mass supported by a stiff u-shaped flexible mount, the cantilever (Fig.~\ref{img:setup}~(a)), following the design of Ref.~\cite{GerberdingMetrologia2015}.
Our sensor features an optical finesse of about two, a resonance frequency of $\omega_0=\SI{678.5}{Hz}$, and a mechanical quality factor of $Q=630$.
Due to its stiffness the optomechanical resonator can be described as an ideal harmonic oscillator.
Below the resonance frequency, displacement of the test mass $X$ as a function of vibration frequency $\omega$ linearly depends on the acting acceleration $A$,
\begin{equation*}
    \frac{X(\omega)}{A(\omega)}=-\frac{1}{\omega_0^2-\omega^2+i\frac{\omega_0}{Q}\omega}\, ,
\end{equation*}
\rev{}{ and is therefore flat. By means of more advanced data analysis, the usable bandwidth can be extended beyond the mechanical resonance.}
Using adhesive bonding, the resonator is attached to a two-inch square mirror retroreflecting the light pulses driving the atom interferometer.
The resonator's acceleration-sensitive axis is aligned collinearly with the retroreflector's normal vector (Fig.~\ref{img:setup}~(a)) by orienting the outer edges of both devices parallel.
The motion of the test mass is read out with a fibre-based optical setup based on telecom components comprising a tunable laser operating at a wavelength near \SI{1560}{nm} protected by an optical isolator~(Fig.~\ref{img:optics}).
The sensor has a quarter wave plate incorporated in its lead fibre thereby enabling us to separate the signal reflected off the resonator using a polarising beam splitter.
Additionally, a small fraction of the laser light is split off before the resonator using a 90:10 splitter.
Making use of differential data acquisition of photo detectors PD 1 and 2 we can therefore cancel common mode laser intensity noise on the resonator signal.
Finally, the processed signal depends on the transmission of the optomechanical resonator and hence the distance between the two reflective surfaces.
Consequently, it is a direct measure of the acting acceleration.
The optomechanical resonator and the retroreflection mirror are operated under normal atmospheric conditions.
We place the mirror with the resonator attached onto a solid aluminium plate in ``strapdown'' configuration on the laboratory floor.
\rev{}{During the initialisation of the optomechanical sensor, a commercial force-balance accelerometer (\textit{Nanometrics Titan}) was utilised to perform test measurements for comparison as well as post correction for reference purposes.}

\rev{}{\paragraph{Data availability --} The data used in this manuscript are available from the corresponding author upon reasonable request.}
\bibliography{main}
\section{Acknowledgements}
We thank H.~Ahlers, J.~Lautier-Gaud, L.~Timmen, J.~M\"uller, S. Herrmann, S.~Sch\"on, K.~Hammerer, D.~R\"atzel, P.~Haslinger, and M.~Aspelmeyer for comments and fruitful discussions and acknowledge financial support from Deutsche Forschungsgemeinschaft (DFG) within CRC 1128~(geo-Q), projects A02, A06, and F01 and CRC 1227~(DQ-mat), project B07, and under Germany’s Excellence Strategy – EXC-2123 QuantumFrontiers – 390837967 (research unit B02).
D.S. acknowledges support by the Federal Ministry of Education and Research (BMBF) through the funding program Photonics Research Germany under contract number 13N14875.
This project is furthermore supported by the German Space Agency~(DLR) with funds provided by the Federal Ministry for Economic Affairs and Energy~(BMWi) due to an enactment of the German Bundestag under Grant No. DLR 50WM1641~(PRIMUS-III), 50WM1137~(QUANTUS-IV-Fallturm), and 50RK1957 (QGYRO), and by ``Nieders\"achsisches Vorab'' through the ``Quantum- and Nano-Metrology~(QUANOMET)'' initiative within the project QT3, and by ``Wege in die Forschung~(II)'' of Leibniz University Hannover.

\section{Author contributions}
W.E.,
E.M.R.,
C.S.,
and D.S. designed the atom interferometer and its laser system.
L.L.R.,
H.A.,
D.N.,
and D.S. contributed to the design of the atom interferometer and its laser system and realised the overall setup.
A.R.,
M.M.,
L.K.,
L.C.,
R.S.,
C.B.,
and F.G. designed, built, and tested the optomechanical resonator and designed the readout laser system.
A.R.,
C.M.,
D.T.,
and \'E.W. built and characterised the laser system for readout.
L.L.R.,
F.G.,
L.K.,
and A.R. implemented the optomechanical resonator in the atom interferometer setup.
L.L.R.,
H.A.,
D.N.,
and A.R. operated the final experimental setup.
Sv.A.,
M.G.,
and C.S. contributed to the data acquisition system utilised for post correction.
A.R.,
D.N.,
C.S.,
and L.L.R. performed the analysis of the data presented in this manuscript.
L.L.R.,
D.S.,
and F.G. drafted the initial manuscript.
A.R.,
C.M.,
C.S.,
D.T.,
\'E.W.,
E.M.R.,
and L.K. provided major input to the manuscript 
and all authors critically reviewed and approved of the final version.
\end{document}